**To Quote this work**

1. Kouakou, P., Brien V., Assouar, B., Hody V., Belmahi M. Migeon H. N., Bougdira J.,
   Preliminary Synthesis of Carbon Nitride Thin Films by N2/CH4 Microwave Plasma Assisted Chemical
   Vapour Deposition: Characterisation of the Discharge and the Obtained Films.
   *Plasma Processes and Polymers* **4**, S210–S214 (2007).
   doi.org/**10.1002/ppap.200730703, hal-02900049**

**Thank you**

# PRELIMINARY SYNTHESIS OF CARBON NITRIDE THIN FILMS BY $N_2/CH_4$ MICROWAVE PLASMA ASSISTED CHEMICAL VAPOR DEPOSITION: CHARACTERIZATION OF THE DISCHARGE AND THE OBTAINED FILMS

Paul Kouakou[1], Valérie Brien[1], Badreddine Assouar[1], Virginie Hody[2], Mohammed Belmahi[1*], Henri N. Migeon[2], Jamal Bougdira[1]

[1] Université de Nancy, Laboratoire de Physique des Milieux Ionisés et Applications (LPMIA), UMR CNRS 7040, Faculté des Sciences et Techniques, boulevard des Aiguillettes B.P. 239; F-54506 Vandoeuvre-lès-Nancy cedex, France

[2] Centre de Recherche Public Gabriel Lippmann, Département Science et Analyse des Materiaux (SAM), 41 rue du Brill-L-4422 Belvaux, Luxembourg

*Corresponding author;
Phone: + 33 (0) 3 83 68 49 24
Fax: + 33 (0) 3 83 68 49 33
E-mail: mohammed.belmahi@lpmi.uhp-nancy.fr

**Abstract**

The present work deals with the synthesis of crystalline carbon nitride thin films by microwave plasma assisted chemical vapour deposition in $N_2/CH_4$ gas mixture. The discharge analysis by optical emission spectroscopy shows that the increase in the $CH_4/N_2$ ratio involves an important production of the CN and $C_2$ radicals. In the films X-ray energy dispersion spectroscopy shows that the N/C ratio decreases when the $CH_4$ percentage in $N_2$ increases. X-ray diffraction and electron diffraction are used to study the carbon nitride films nature. Scanning electron microscopy shows that the films consisted of nano-crystalline grains. Carbon balls are also present on the film surface for $CH_4$ percentage higher than 4%. The transmission electron microscopy confirms the nano-structure of the film and shows the isotropic etching of the substrates, during the film growth.

# 1. Introduction

In 1989 Liu and Cohen predicted that the $\beta$-$C_3N_4$ would exhibit properties close to those of diamond,[1] enabling many applications, for instance in mechanics and medicine.[2] Many deposition techniques were used to attempt the carbon nitride materials synthesis, namely: ion beam deposition techniques or plasma based techniques such as pulsed laser deposition (PLD),[3-4] electron cyclotron resonance deposition (ECR),[5-6] direct current (DC) or radio frequency (RF) deposition,[7-8] inductively coupled plasma chemical vapour deposition (ICP-CVD),[9] microwave plasma assisted chemical vapour deposition (MPACVD).[10-12]

Several reviews comparing these techniques of synthesis and the characterization of the resulting films were published.[13-15] They make critical comments on the advantages and drawbacks of the elaboration techniques. Indeed, doubt still exists whether $\beta$-$C_3N_4$ can be prepared or not, insofar as nobody has reached the synthesis of pure $\beta$-$C_3N_4$. At best the crystalline $\beta$-$C_3N_4$ phase was obtained in very small proportion and was always embedded in phases like the $\alpha$-$C_3N_4$, cubic $C_3N_4$, graphitic $C_3N_4$ or amorphous phase.

MPACVD is a very powerful technique, which was successfully used to synthesise diamond using $H_2/CH_4$ gas mixture. Unfortunately, polycrystalline diamond deposits do not meet all industrial requirements for mechanical applications because of fragility and delamination problems. There is still good hope that $CN_x$ polycrystalline films, which are probably at least as hard as Carbon diamond ($C_d$) will solve these problems. MPACVD gave excellent results for good quality $C_d$ synthesis, and was naturally tested for the $CN_x$ ones. First results obtained are still not fully satisfying. It is then so proposed to improve the MPACVD process by understanding the growth mechanisms. To do so, special care is taken to correlate the plasma diagnostic data to the films characteristics. The growth was carried out using Si substrates and $N_2/CH_4$ gas mixture. Firstly, systematic study is performed to determine the influence of the $CH_4$ %. Secondly, to precise the growth mechanism, the optimised condition found on silicon substrates was exploited on a $Si_3N_4$/Si substrate.

## 2. Experimental

The growth of carbon nitride films is carried out in a MPACVD reactor which was described elsewhere.[16] The gas mixture is composed by $N_2$ and $CH_4$ in various ratios. Films are initially deposited on Si (100) substrates and thereafter on Si/$Si_3N_4$ substrates. Prior to deposition, substrates are ultrasonically cleaned with various solvents (methanol, trichloroethylene, deionised water).

After introduction of the substrate into the chamber, the treatment starts when the residual pressure is about $10^{-5}$ Pa. The working total pressure inside the reactor lies in the range of 3-10 kPa. The typical total gas flow ranges from 30 to 150 sccm. The microwave power used can vary between 600 and 2000 W. The apparent temperature of the sample, measured with a monochromatic ($\lambda$=1.6 µm) infrared pyrometer evolves between 700 and 1200 °C during the treatment.

The characterization of the plasma is made by OES using a Jobin Yvon TRIAX 550 monochromator equipped with a CCD camera. The emitted light of the plasma is collected through an optical fibre. The observed spectral region extends from 300 to 700 nm. This technique will allow correlations between the plasma chemistry and the film properties.

Scanning electron microscopy (SEM) observations are performed on a JEOL JSM 6500 F using an accelerating voltage of 5 kV to analyse the general morphology of the films.

The transmission electron microscopy (TEM) is used to deepen the knowledge on the films morphology. TEM observations are carried out on a Philips CM20 microscope operating at an accelerating voltage of 200 kV. They are performed on matter chips sampled on the films by the microcleavage technique.

Chemical composition of the films is systematically estimated by energy dispersive X-ray spectroscopy (EDSX). The EDSX spectra are recorded on the CM 20 Philips microscope equipped with an ultra thin window X-ray detector. Some samples are also analyzed by Scanning Auger Electron Spectroscopy (AES) in order to determine their chemical composition, which will be given in atomic percentage.

The crystallographic structure is determined by X-ray diffraction. The θ−2θ XRD patterns are performed on a Philips X'PERT PRO apparatus (Cu K$_\alpha$ radiation, λ=1.54056 Å).

### 3. Results and discussion

#### 3.1 Optical emission spectroscopy

The optical emission spectroscopy (OES) is a non-intrusive technique used to identify the emissive species present in the plasma and to follow their density evolution while varying the experimental parameters. Indeed, the intensity emitted by specie is proportional to its density.[17]

Figure 1a shows a typical spectrum recorded for $N_2$/$CH_4$ plasma. Such a spectrum presents the emission bands of CN, $C_2$ and $N_2$ species. CN and $C_2$ radicals are known to play an important role during carbon nitride growth.[18-21] The evolution of the intensities of CN violet system [$B^2\Sigma^+ \rightarrow X^2\Sigma^+$ ($\Delta v=0$)] at 388.34 nm and $C_2$ swan system [$A^3\Pi_g \rightarrow X^3\Pi_u$ ($\Delta v=0$)] at 516.52 nm were followed as a function of the experimental parameters.

The $N_2$/$CH_4$ plasmas in MPACVD system are stable in a small interval of microwave power and total pressure in our study. The useful parameters permitting the achievement of a stable MPACVD discharge shape suitable for large sample surface treatments, important duration and various $CH_4$ percentages have been determined. These parameters were thus fixed to 1000W for the microwave power, 5000 Pa for the total pressure and 50 sccm for the total flow rate.

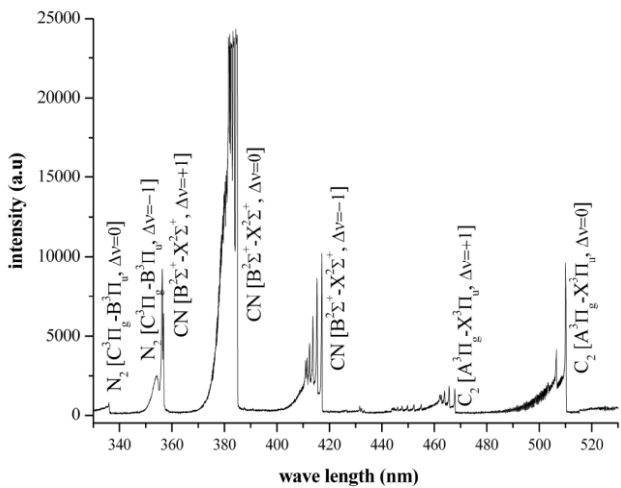

fig 1a

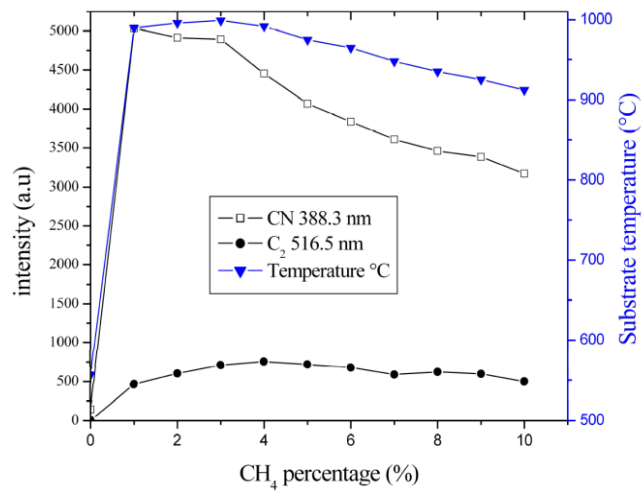

fig 1b

Figure 1   (a):Typical OES spectrum obtained for $N_2$/$CH_4$ gas mixture in MPACVD system. The experimental parameters are: 1000 W, 2% $CH_4$, 5 kPa, 50 sccm.
   (b):   Evolution of the CN, $C_2$ emissions and the substrate temperature 5 mm above the surface as a function of the $CH_4$ percentage in nitrogen. (1000 W, 5 kPa, 50 sccm).

Figure 1b shows the evolution of the CN and $C_2$ emissions versus the percentage of $CH_4$ in gaseous phase. The CN emission intensity and the temperature increase abruptly when a small quantity of $CH_4$ is injected. These two parameters then progressively decrease to reach a value 30 % lower for a $CH_4$ percentage of 10 %. This can be attributed to the drop of nitrogen percentage and/or the displacement of the plasma relatively to the substrate surface.

The intensity of the $C_2$ emission increases with the $CH_4$ percentage until 4% and then slightly decreases over the rest of the range. The surface temperature curve profile is similar to that of the CN emission intensity. The follow-up these two emission evolutions thus showed that the highest reactivity of the MPACVD $N_2$/$CH_4$ discharge is obtained for $CH_4$ percentage lower than 4%.

### 3.2 Films characterization

Several techniques of characterization were combined to determine the morphology and the chemical composition of the films.

TEM investigations (figure 3) confirm the SEM micrographs of the sample: the films are polycrystalline. For a $CH_4$ percentage smaller than 4%, the grain shape looks rounded, whereas above this value; the grains clearly exhibit abrupt angles.

The TEM cross-section view of the film (Figure 3b) shows that, the film is approximately 60 nm thick.

The spherical voids observed at the substrate/film interface clearly show that the erosion is isotropic proving the chemical nature of this process during the growth.

The voids observed at the substrate/film interface are spherical, such an etching geometry is typical from chemical etching processes. If the process had been physical etching, the obtained geometry should have been anisotropic. This behaviour can only be observed when the substrate is negatively biased, favouring ionic bombardment.

Auger measurements revealed that the balls are only made of carbon and that the underlying film contains N, C, Si and O. The chemical composition of the underlying layer was also analysed by EDSX and the N/C ratio was calculated. All the results are summarized in table I. These analyses inform that the films are very rich in silicon.

However, the silicon concentration is certainly overestimated. As a matter of fact, the way the measurements are done implies that a small part of silicon coming from the substrate is probably analyzed with the film. Moreover, this shift to high values is reinforced by a higher sensitivity of the technique for Si than for N, C and O.

The crystallographic characterization performed by $\theta-2\theta$ X-ray diffraction on the films showed that they are all similar. One pattern is exemplified in figure 4. The atomic distances ($d_{hkl}$) were compared to the ones of the JPCDS database (n° 50-1250 for $\beta$-$C_3N_4$, n° 1-078-1693 for cubic $C_3N_4$ and n° 1-074-2309 for SiCN). Some of the peaks could be identified and could be attributed to either the $\beta$-$C_3N_4$, cubic $C_3N_4$ or the SiCN phase. However some experimental peaks were not identified. In addition, all peaks of the $\beta$-$C_3N_4$, cubic $C_3N_4$ and the SiCN phase do not appear on the diffraction patterns.

Such observation indicates that the identification of the phases requires further investigations. Electronic diffraction characterizations confirmed this point.

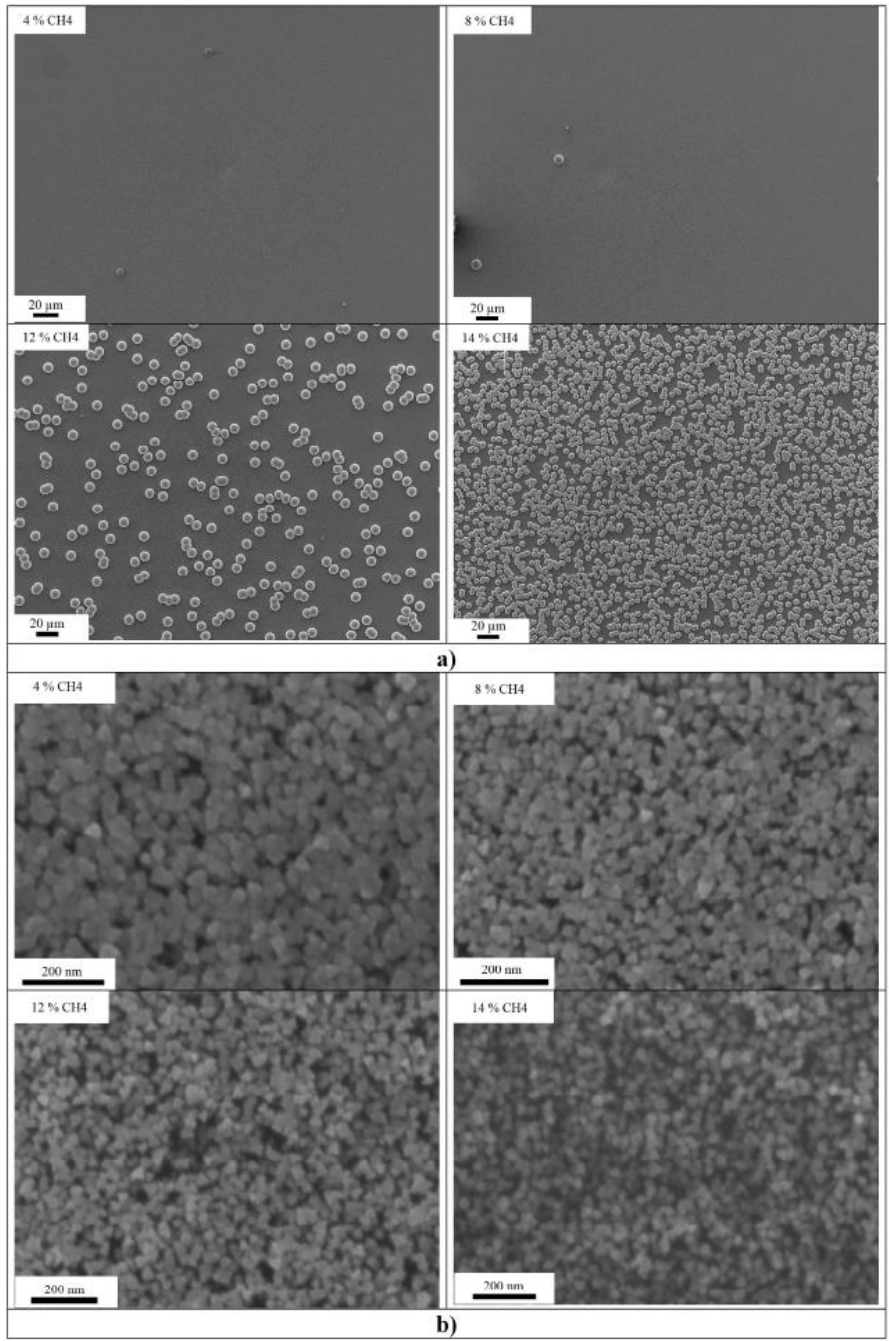

Fig 2

Figure 2: SEM micrographs of carbon nitride films for different $CH_4$ percentages
(a) Low magnification, (b) high magnification

Figure 2 presents low and high magnification SEM micrographs carried out on the films obtained with different $CH_4$ percentage. Films are made of crystalline grains and exhibit balls on the surface at high $CH_4$ percentages. The balls density increases and the average grain size (50 nm at 4 % $CH_4$, 30 nm at 12 % $CH_4$) slightly decreases when the $CH_4$ percentage increases.

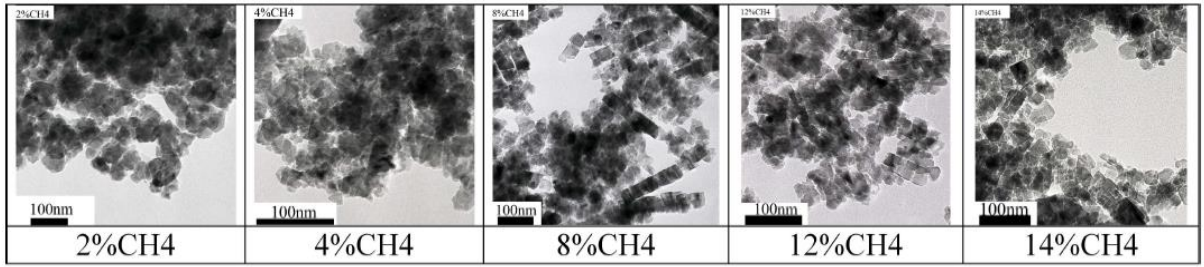

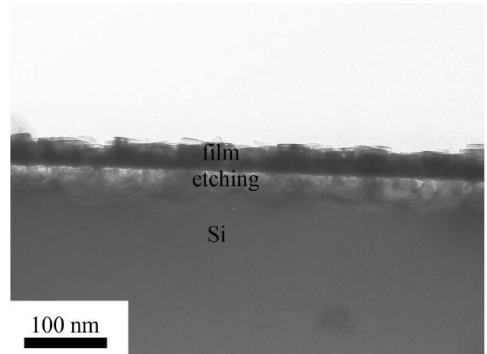

Figure 3: TEM micrographs of carbon nitride films
(a) For different $CH_4$ percentage, (b) cross-section view of the film

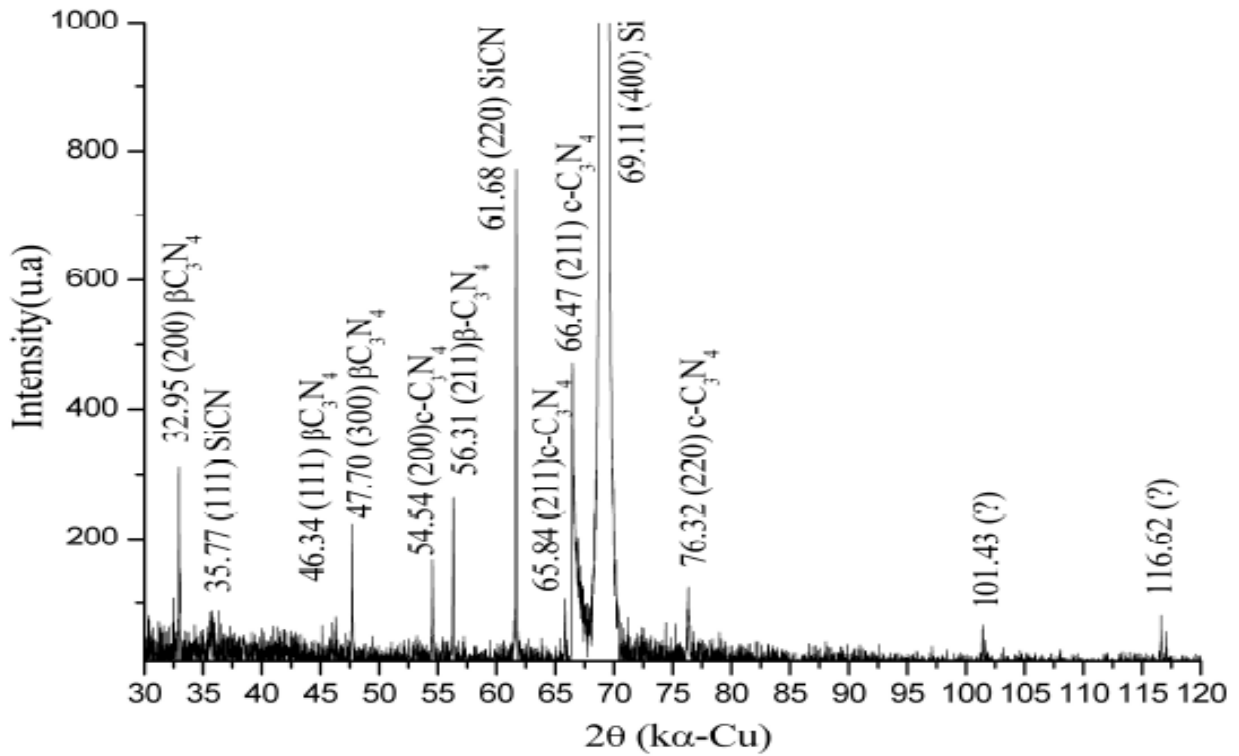

Figure 4: Experimental X-ray diffraction of typical carbon nitride film (2% $CH_4$, 1000 W, 50 sccm, 50 mbar) deposited on Si compared with JCPDS database
(file 50-1250 for $\beta$-C3N4 , 1-074-2309 for SiCN and 1-078-1693 for cubic $C_3N_4$)

### 3.3 Discussion

As shown by the XRD electron diffraction experiments, the phases present in the films are limited to $\beta$-$C_3N_4$, cubic $C_3N_4$ and SiCN. Probably other phases are present in the films depending on the experimental conditions, as actually observed in literature.[22] To avoid carbon balls formation on the films surface, it is necessary to use a $CH_4$ percentage lower than 4 %.

At these values, the films are still too rich in silicon, and too poor in nitrogen (N/C (0.5-0.7)). Indeed the $\beta$-$C_3N_4$ N/C ratio is 1.33.

According to the results published by Jiang *et al* the N/C ratio obtained at low temperature (100-700°C) is about 50% and the obtained films are mainly amorphous.[23] In the temperature range of 780-900°C, Zhang *et al* obtained crystalline films with a N/(N+C) ratio varying as a function of the temperature between 0.6 and 1.2.[24] Special care must be taken when examining this results because the emissivity of the film surface varies during the growth, thus the temperature.

Obviously, high temperature is not sufficient to increase nitrogen incorporation in the film. In our case, we work at high temperature in order to synthesize crystalline $CN_x$ films. To insert more nitrogen inside the films, we propose a preliminary growth on another type of substrate, namely Si/$Si_3N_4$ (100 nm of $Si_3N_4$ on Si). The chosen experimental conditions for this deposition were 1000 W, 50 sccm, 4 % $CH_4$, 5000 Pa. The TEM cross-section view displayed in figure 5a shows that neither the morphology of the films (compare with figure 3b) nor the nature of the etching are affected by the substrate nature modification. The EDSX results give N: 28 %, C: 31%, O: 2% and Si 39%. The nitrogen content has then been slightly improved (N/C=0.9).

A possible growth mechanism can be proposed. First, two kinds of sites have to be considered: germination and etching sites. On germination site, carbon, nitrogen and silicon recombine and deposit to growth the film, at the contrary, on etching site species coming from the plasma (atomic nitrogen, atomic hydrogen …) desorbs the adsorbed surface elements ( Si, C…) that are further pumped. This is principally the cause of the substrate etching. It has already been mentioned that the etching seems to be chemical and the etching element is mainly atomic nitrogen. The etching produces, CN and probably SiN radicals and the growth simultaneously occurs on the germination sites. The EDSX profile presented in figure 5b shows that the silicon content decreases with the increasing of the film thickness, which never exceeds 100 nm even for high deposition times.

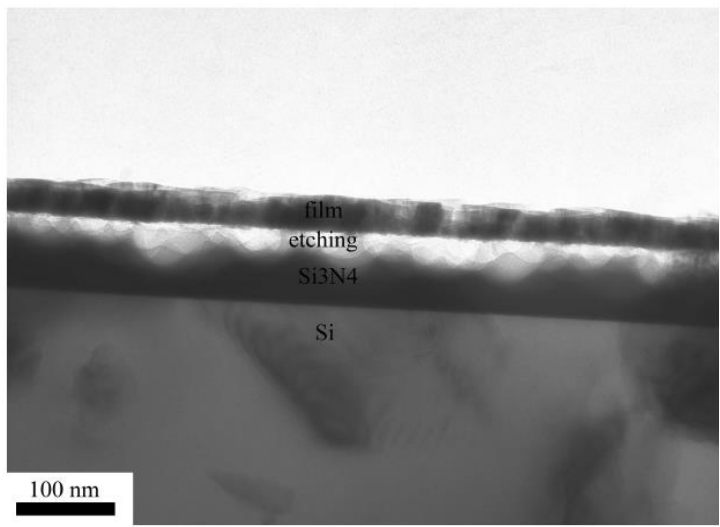
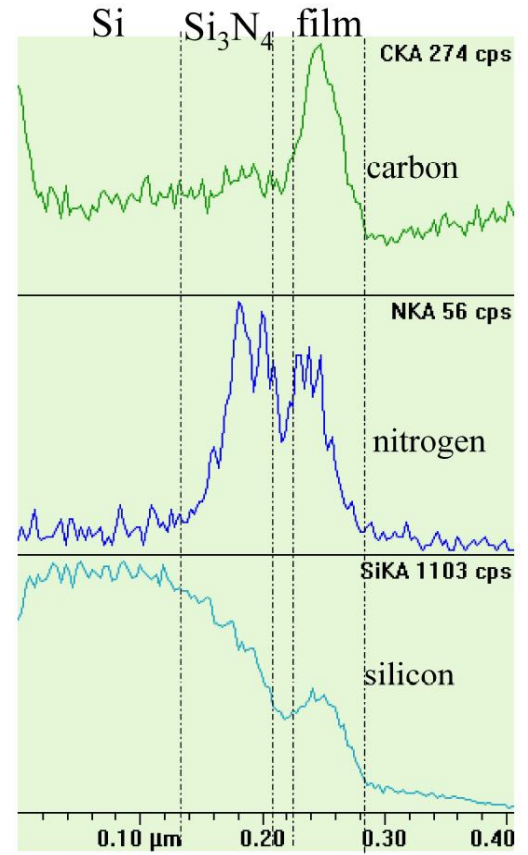

Figure 5: (a) TEM cross view of film on Si/Si$_3$N$_4$ substrate, (b) EDSX profile

## 4. Conclusion

Nano-crystalline carbon nitride films were synthesised by MPACVD in CH$_4$ and N$_2$ gas mixture. The OES analyses of the plasma show that the highest reactivity of the discharge is obtained for a CH$_4$ percentage lower than 4%. The study of the morphology and the nature of the films as a function of the CH$_4$ percentage shows that their thickness is limited to approximately 100 nm. This study also highlights that the growth is assisted by the etching of the silicon. Further experiments are needed to clarify the exact nature of the films.

**Acknowledgments**

This project was supported by the Fonds National de la Recherche of Luxembourg. This work is also partly supported by the Ministère de la Culture, de l'Enseignement Supérieur et de la Recherche of Luxembourg, by the Centre de Recherche Public Gabriel Lippmann of Luxembourg and by the Ministère délégué à l'Enseignement Supérieur et à la Recherche of France. The authors wish to thank J. Ghanbaja, and J. P. Emeraux for TEM and X-ray analyses.

# Table captions

| %CH$_4$ | [N](%) | [C](%) | [O](%) | [Si](%) | N/C | N/(N+C) |
|---|---|---|---|---|---|---|
| 2 | 20 | 30 | 5 | 45 | 0.67 | 0.4 |
| 4 | 15 | 30 | 6 | 48 | 0.5 | 0.33 |
| 8 | 10 | 40 | 5 | 44 | 0.25 | 0.20 |
| 12 | 12 | 49 | 6 | 33 | 0.24 | 0.20 |
| 14 | 10 | 65 | 4 | 21 | 0.15 | 0.13 |

Table1: Atomic concentrations and N/C ratio of the carbon nitride films deposited on Si for different CH$_4$ percentages

# Journal´s table of contents:

Carbon nitride thin films were synthesised by MPACVD in N$_2$/CH$_4$ gas mixture. OES was used for *in situ* diagnostics of the plasma and the films were characterized by SEM, TEM and DRX. Films were nano-crystalline and may contain *β*-C$_3$N$_4$, *cubic*-C$_3$N$_4$ and SiCN. A growth mechanism was proposed.